# SED models of AGN

R. Siebenmorgen[*] and A. Efstathiou[†]

[*]*ESO, Karl-Schwarzschildstr.2, D85748 Garching b.M., Germany homepage*
[†]*School of Computer Science & Engineering, Cyprus College, 6 Diogenes Street, Engomi, 1516 Nicosia, Cyprus*

**Abstract.**
   Present AGN models aiming to account for the observed infrared SEDs consider a physical description of the dust and a solution of the radiative transfer problem. Mid infrared spectra obtained at different spatial scales are presented. They show that PAH bands are detected in starburst regions but significantly reduced near the centre of AGN. This may be explained by examining the heating mechanism of PAHs after hard photon interactions. On the radiative transfer side first a most economic model is presented where three parameters, luminosity, effective size and extinction of the nucleus are varied to obtain SED fits. A full grid of model spectra are made available at: http://www.eso.org/~rsiebenm/agn_models/. This model is sufficient to account for ISO broad band data of a sample of 68 radio galaxies and quasars of the 3CR catalogue. The hot dust component detected is mainly due to small grains and PAHs. In such models, type 1 AGNs are represented by a compact dust distribution with warm grains and weak PAH emission. In AGNs of type 2, the dust appears to be more extended, relatively colder and PAH bands are strong. Realistic AGN models which are consistent with the unification need to explain the overall absence of the 9.7$\mu$m silicate emission feature. This can be done by considering various geometries (tapered discs). Models which combine AGN and starburst activity are presented for galaxies with hidden broad line region. It is found that the AGN torus dominate the mid IR continuum emission and that the starbursts dominate the PAH band as well as the far infrared and submillimeter emission.

## 1. INTRODUCTION

Nuclei of luminous infrared galaxies are generally dust enshrouded and not transparent. Therefore radiative transfer calculations have to be carried out for an optically thick dusty medium. This has been done in various approximations for starburst and AGN dominated galaxies (Krügel & Tutokov 1978, Rowan-Robinson & Crawford 1989, Pier & Krolik 1993, Loar & Draine 1993, Granato & Danese 1994, Krügel & Siebenmorgen 1984, Siebenmorgen et al. 1997, Silva et al. 1998, Efstathiou et al. 2000, Ruiz et al. 2001, Siebenmorgen et al. 2001, Nenkova et al. 2002, Takagi et al. 2003, Siebenmorgen et al. 2004a,b) and in this conference by Dopita et al., van Bemmel et al. and Arimoto et al. All these groups aim in a physical model of the infrared emission of galaxies. Others apply an empirical approach and fit SED by templates of known objects. For pitfalls of such empirical methods see Dole et al.

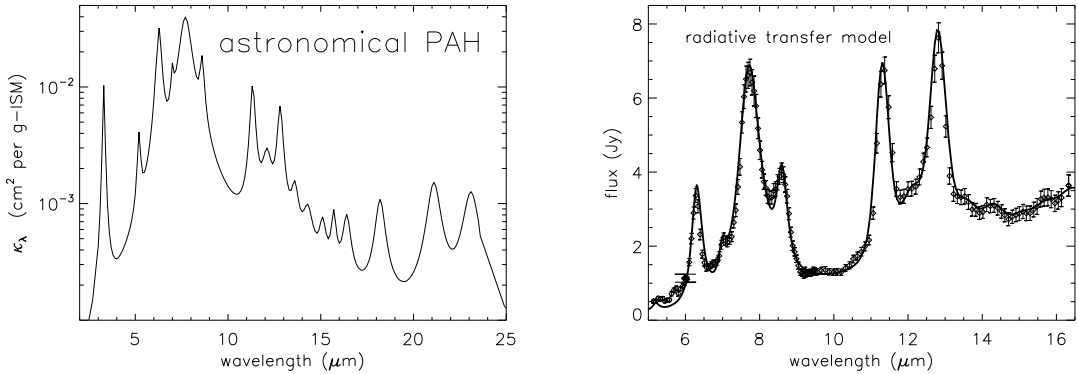

**FIGURE 1.** *a (left):* Infrared absorption coefficient $\kappa_\lambda$ of astronomical PAH. There are two populations, one with $N_C = 30$ C atoms, the other with $N_C = 500$. Both populations have an hydrogenation parameter $f_{H/C}=0.2$ and an abundance such that for $10^5$ protons in the gas phase one finds one carbon atom in each PAH population. *b (right):* Comparison of the radiative transfer model after convolution with the ISOCAM CVF spectral response function (full line) with the ISOCAM spectrum of NGC1808.

## 2. THE DUST MODEL

The dust consists of large carbon and silicate grains with radii, $a$, between 300 and 2400Å and a size distribution $n(a) \propto a^{-3.5}$; very small graphites ($a = 10$Å); and two kinds of PAHs (30 C and 20 H atoms; 252 C and 48 H atoms and an hydrogenation parameter $f_{H/C}=0.2$. The PAH abundance is such that for $10^5$ protons in the gas phase one finds one carbon atom in each PAH population. The infrared absorption coefficient $\kappa_\lambda$ of astronomical PAH is derived by Siebenmorgen et al. (2004a) and shown in Fig. 1a.

When a grain absorbs an optical or UV photon of frequency $\nu$, its internal energy rises by the full amount $h\nu$. Furthermore, when the grain is not smaller than the wavelength, the absorption efficiency is approximately one. The situation is different at X–rays. The interaction of hard photons with interstellar dust and the limits of Mie theory have been studied by Voit (1991), Laor & Draine (1993) and Dwek & Smith (1996). Below 10 keV, the main absorption process is photo–ionization of inner atomic shells. The excited electron loses energy in inelastic scattering as it travels through the grain, but may also leave the grain and carry away kinetic energy as a photo–electron. When an X–ray photon has created a gap in the innermost shell, the gap will be filled by a downward transition of an electron from an upper shell and the energy may escape either as a photon, or non–radiatively through ejection of an Auger electron. The computations of Dwek & Smith (1996) indicate that for graphite particles of 50Å radius, soft X–ray photons with $h\nu \lesssim 100$ eV deposit practically all their energy in the grain and Mie theory is still valid. For photon energies above $h\nu = 100$ eV and particle radii $a < 50$Å, only part of the photon energy is deposited in the grain and we therefore reduce the absorption efficiency by a factor $\propto 1/\nu$.

The dust model presented here together with a radiative transfer computation is sufficient to fit SEDs which are observed with ISO (Siebenmorgen et al. 1999). In Fig.1b we give one example in which our starburst model is compared to the ISOCAM spectrum

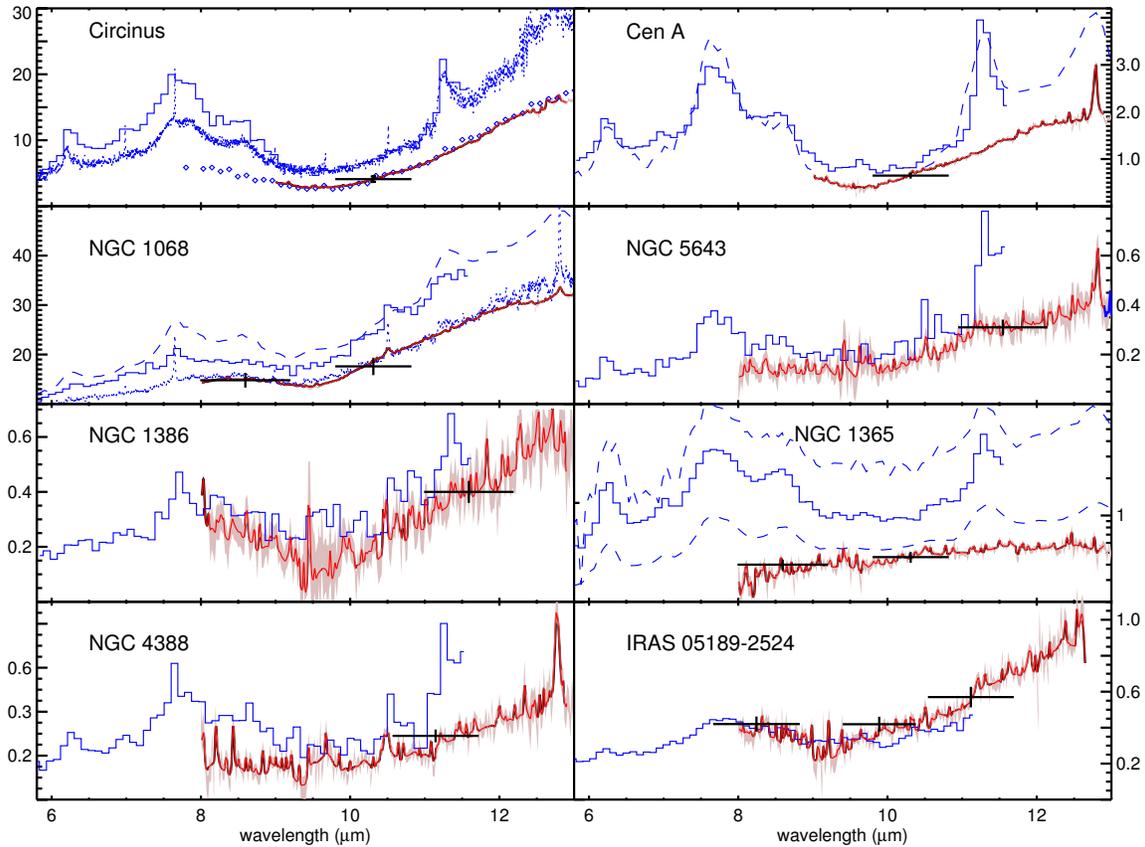

**FIGURE 2.** Mid-infrared flux densities in Jy of Seyfert galaxies (Siebenmorgen et al. 2004a): histograms and dashed lines are large beam ($14'' - 20''$) ISO observations, thick lines are narrow slit ($3''$) data with TIMMI2. Note that PAH bands are visible in the large beam spectra but absent in the high spatial resolution data of the same galaxy.

of NGC1808 (Siebenmorgen et al. 2001).

## 3. PAH EVAPORATION

Active galaxies with Seyfert nuclei often show strong PAH emission bands when observed at large scales, say with the spatial resolution of ISO ($\geq 10''$). However mid infrared spectra of the same galaxy are PAH free when observed at high spatial resolution and where the slit is centered close to the nucleus. A few examples of this finding are presented in Fig.2. This observational fact and a study of the survival of PAHs in starburst and AGN environments with energetic photons of, at least, a few up to hundreds of eV is presented by Siebenmorgen et al. (2004a). They found that small grains evaporate

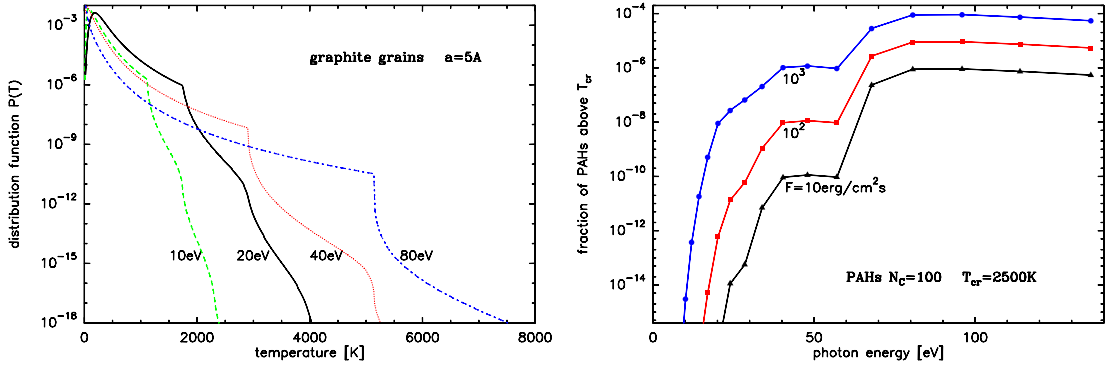

**FIGURE 3.** *a (left):* The distribution function of the temperature, $P(T)$, for graphite grains of 5Å radius. Monochromatic flux $F = 10^4$ erg cm$^{-2}$ s$^{-1}$, photon energies from 10 to 80eV. *b (right):* The fraction of PAHs with $N_C = 100$ carbon atoms above a above a critical temperature $T_{cr} = 2500$ K. The strength of the monochromatic flux $F$ is indicated.

after interaction with hard photons and in particular with photons of more than 50eV which can only be emitted by X–ray sources. A simple explanation is found by studying the quantum statistical behaviour of PAHs:

Let $P(T)$ be the temperature distribution function (Siebenmorgen et al. 1992) of an ensemble of very small grains, for example, PAHs, so that $P(T)\,dT$ gives the probability of finding a particular grain in the temperature interval $dT$ around $T$. Whereas grain cooling usually proceeds through emission of infrared photons, in a very hot particle, above $\sim 2500$ K, evaporation is more effective. The distributions $P(T)$ in Fig. 3a display a kink at the temperature where the thermal energy of the grain equals $h\nu$. Still higher grain (momentary) temperatures are achieved when a second photon is absorbed before the grain has cooled off from the excitation by the first. Let us assume that the grains evaporate if a certain fraction of them, $f_{ev}$, are above some critical temperature $T_{cr}$. The latter should be somewhat greater than $T_{ev}$. Relegating all complicated or unknown physics to the parameters $f_{ev}$ and $T_{cr}$, we write the condition for evaporation as

$$I_{ev} \equiv \int_{T_{cr}}^{\infty} P(T)\,dT \; > \; f_{ev}\,. \tag{1}$$

Reasonable numbers for $T_{cr}$ in (1) are 2500 K and $f_{ev} \sim 10^{-8}$, but such suggestions do, of course, not constitute a theory.

Fig. 3b displays the integral $I_{ev}$ in equation (1) as a function of photon energy $h\nu$ for various fluxes and PAH with $N_C = 100$ carbon atoms. The important point is here that for the smallest graphite grains with 5Å radius (65 carbon atoms), $I_{ev}$ grows very rapidly in the range from 20 to 40eV and then declines only gradually. The energy of photons, $\langle h\nu \rangle$, emitted by the hottest O stars ($T_{eff} = 5 \cdot 10^4$ K) is only 16 eV when averaged over the total spectrum, and equals 22 eV when averaged over the Lyman continuum ($\lambda < 912$Å). These numbers are thus upper limits for the radiation field in a starburst. Furthermore, the stellar flux of an O star plummets at wavelengths below 228Å, or above $h\nu = 54.1$ eV, when helium becomes doubly ionized. AGNs, on the other hand, have a spectrum typically declining with $\nu^{-0.7}$ and emit copious amounts of extreme UV and soft X–ray photons as witnessed by the observation of fine structure lines from ions with high ($> 100$ eV) ionization potentials.

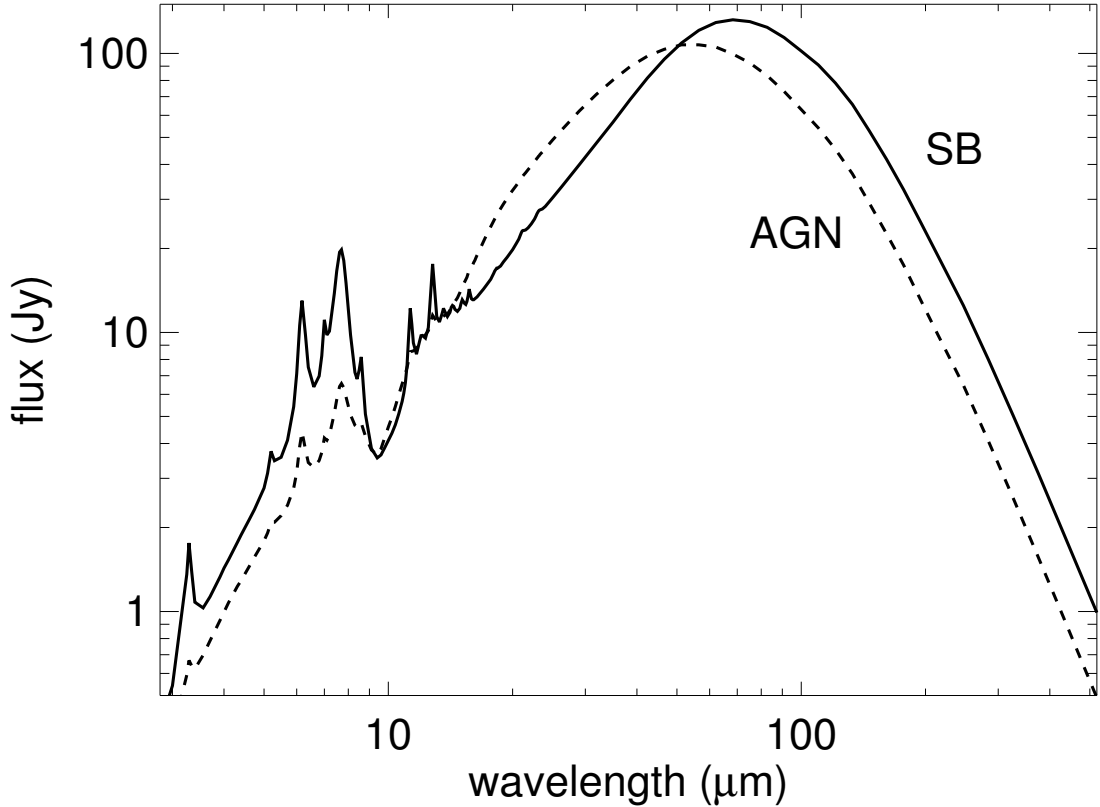

**FIGURE 4.** Comparison of radiative transfer model spectra of dusty starbursts (solid) and AGN (dashed). In both models the visual extinction towards the cloud centre is 25mag and the luminosity is $L = 10^{11} L_\odot$. Spectra are calculated for a distance of 50 Mpc.

We conclude from Fig. 3 that small graphite grains or PAHs (5Å radius) survive in a starburst as long as fluxes are below $\sim 10^4$ erg cm$^{-2}$, but they will be destroyed near an AGN of the same flux $F$. The fate of PAHs of equivalent mass is similar, larger grains are much more resistive. They need higher fluxes to get vaporized and survive in a starburst environment.

## 4. RADIATIVE TRANSFER MODELS

Here we present comprehensive model calculations for the mid–infrared emission of starburst and AGN radiation environments in the optically thick regime.

The radiative transfer in a dust cloud around a starburst is computed following Siebenmorgen et al. (2001), around an AGN after Siebenmorgen et al. (2004b). The major difference between both radiative transfer models is that AGN are heated by a central source whereas starburst galaxies are heated by stars which are distributed over a large volume. In case of a starburst, each of the OB stars have a uniform luminosity of $20\,000 L_\odot$ and a stellar temperature $T_* = 25\,000$ K. The stellar density changes with galactic radius $r$ up to 500 pc from the center like $\propto 1/r$. Each OB star is surrounded by a *hot spot* of constant dust density, corresponding to $n(\mathrm{H}) \sim 10^4 \,\mathrm{cm}^{-3}$. As for the AGN models, the inner radius of the hot spots is given by the photo–destruction or evaporation radius of the grains. The outer radius of the hot spot is determined by the condition of equal heating of the dust from the star and from the radiation field in the galactic nucleus. Therefore for starburst galaxies the radiative transfer has to be solved for two different scales: first, on galactic scale of the nucleus and second, for each of the hot spots. Both transfer problems are linked with each other via the boundary conditions (for example the size of the hot spots) and an iterative scheme is applied to find a self consistent solution. Starburst models presented by other groups often simplify matters and consider the first problem for ensemble of hot spots or molecular clouds in an optically thin galactic medium. A full grid of starburst model spectra with a large variation of key parameters (as done for the AGN models described in Sect. 5) is available after request.

In Fig. 4. we present our model spectrum for a starburst and an AGN. Both models are only distinguished by the primary heating source of the dust (stars or AGN); otherwise all model parameters such as luminosity ($10^{11} L_\odot$), dust density distribution, total extinction (25mag) and dust model (Sect. 2) are identical. We find that the starburst shows strong PAH emission, whereas for AGN the PAH bands are $\sim 5$ times weaker. Also in the far IR the starbursts are much cooler and dominate the emission when compared to the AGN.

## 5. RADIO-LOUD AGN AND QUASARS

A survey of all 3CR sources imaged with ISOCAM and ISOPHT was recently published by Siebenmorgen et al. (2004b) and Haas et al. (2004), respectively. The sample consists mostly of radio–loud active galactic nuclei (AGN). In total, 68 objects of the 3CR catalogue are detected. In order to discuss properties of the dust emission we apply new radiative transfer models. In the models the dust is heated by a central source which emits photons up to energies of 1keV. As the data coverage in the infrared is sparse we had to develop an economic model with a minimum of free parameters. We use three parameters: luminosity, effective size and extinction. Reasonable fits to the SED of 3CR sources are found for objects where the infrared is dominated by dust re-emission. One of such fits is presented in Fig. 6.

From the sample we (Freudling et al. 2003) study differences of dust signatures by selecting four different subsamples: high ($> 10^{11} \, L_\odot$) and low ($\leq 10^{11} \, L_\odot$) luminosity sources and type 1 and 2 AGNs are distinguished. The mean of the observed fluxes and composite model spectra are shown in Fig 5 for each of the different object classes. In the models, type 1 AGN can be represented by a compact dust distribution, the dust is therefore very warm and emission of PAHs is weak because of photo–destruction (Sect. 3) and the much stronger underlying continuum emission by the large grains. In AGNs of type 2, the dust appears to be relatively colder but PAH bands are strong. Our models contain also dust at large (several kpc) distance from the AGN. Such a cold dust component was neglected in previous computations which therefore underestimated the AGN contribution to the far infrared (FIR).

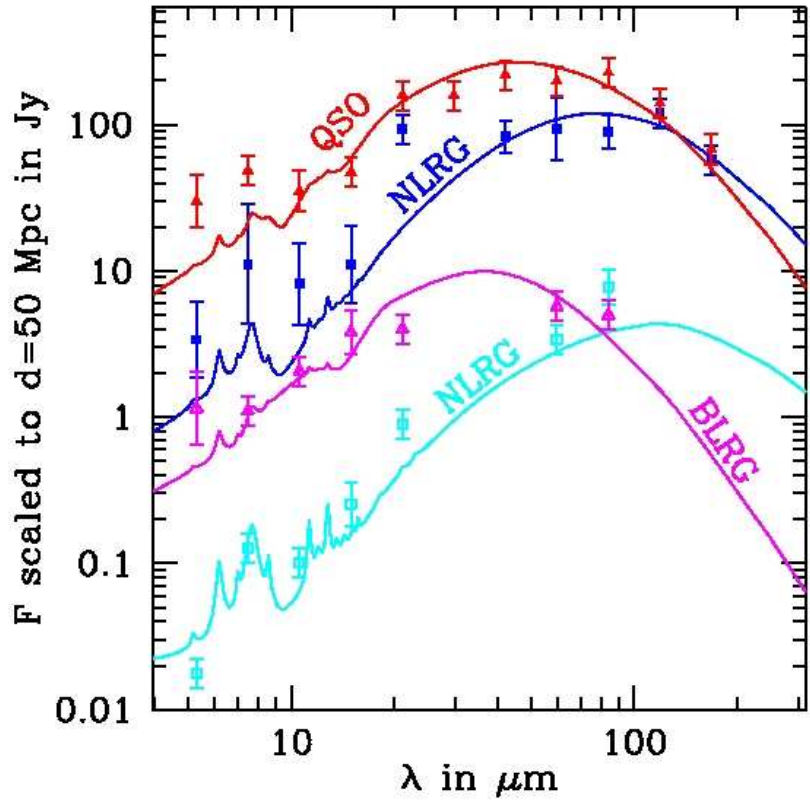

**FIGURE 5.** Mean spectral energy distribution from 53 sources sorted by AGN type and two luminosity bins (Freudling et al. 2003, Siebenmorgen et al. 2004b). Fluxes in Jy are scaled to a distance of 50Mpc. Symbols represent logarithmic averaged flux in wavelength band. The $1\sigma$ error bar is indicated. Solid lines show the corresponding composite model for each AGN class.

The full grid of AGN model spectra ranging from $10^9$ to $10^{14}$ $L_\odot$, visual extinction between 1 to 128mag and outer radii from 125pc to 16000pc is available at: http://www.eso.org/~rsiebenm/agn_models/.

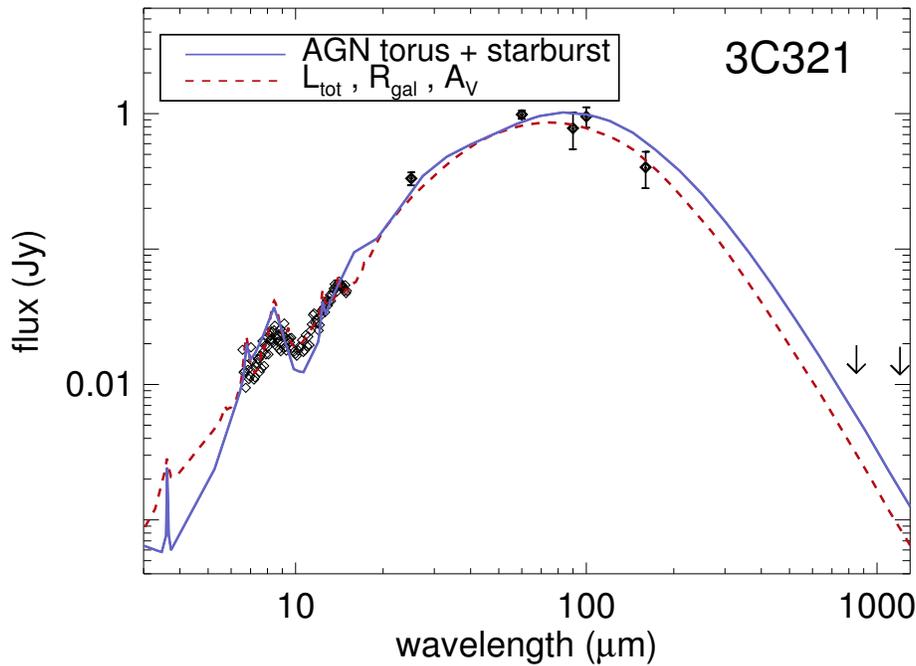

**FIGURE 6.** Spectral energy distribution of 3C321. ISO data are from Efstathiou & Siebenmorgen (2005). We compare two radiative transfer models: a) a three parameter fit (dashed line) using the AGN model by Siebenmorgen et al. (2004b) and a combination of a tapered torus model with starburst activity by (Efstathiou & Siebenmorgen 2005). Also both models are consistent with the observations the later is in better agreement with the unification model and provides correction for the anisotropic torus emission.

## 6. AGN AND STARBURST COMPOSITE SPECTRA

Most active galaxies are known to be composite objects where a Seyfert nucleus is surrounded by a ring of star formation. For NGC1068 the AGN torus in the center should be a few dozens of pc (Siebenmorgen et al. 1994) while the starburst ring has a diameter of a few kpc (Telesco et al. 1984). Here we apply starburst models by Efstathiou et al. (2000) and two dimensional tapered disc models for the AGN by Efstathiou & Rowan–Robinson (1995). The models are discussed in more detail by Efstathiou & Siebenmorgen (2005) who apply them to fit ISO data of galaxies with hidden broad line regions. In total six free parameters are used. One still preliminary example of the model fit is given for 3C321 in Fig.6. Also the fit seems to be as good as the one presented by a three parameter solution of the models presented in Sect.4 it has several advantages. The model presented here enable us to separate the contribution from the dusty disc of the AGN and the dusty starbursts. We find that tapered discs dominate the emission in the mid infrared part of the spectrum and the starbursts in the far infrared. The AGN models by Efstathiou have a two dimensional structure therefore they provide a correction factor of the AGN luminosity for anisotropic emission. In this sense they are more consistent with the unified model.